\documentclass[12pt]{article}
\usepackage{amssymb,amscd,amsmath, array}
\usepackage[active]{srcltx}
\catcode `\@=11
\@addtoreset{equation}{section}

\newtheorem{thm}{Theorem}[section]

\def\qed{\blacksquare}
\newcommand{\be}{\begin{equation}}
\newcommand{\ee}{\end{equation}}
\newcommand{\bea}{\begin{eqnarray}}
\newcommand{\eea}{\end{eqnarray}}
\newcommand{\R}{\mathbb{R}}
\newcommand{\N}{\mathbb{N}}

\begin{document}
\begin{titlepage}

\begin{center}
{\bf \Large{The Higgs Sector in the Causal Approach\\}}
\end{center}
\vskip 1.0truecm
\centerline{D. R. Grigore, 
\footnote{e-mail: grigore@theory.nipne.ro}}
\vskip5mm
\centerline{Department of Theoretical Physics, Institute for Physics and Nuclear
Engineering ``Horia Hulubei"}
\centerline{Institute of Atomic Physics}
\centerline{Bucharest-M\u agurele, P. O. Box MG 6, ROM\^ANIA}

\vskip 2cm
\bigskip \nopagebreak
\begin{abstract}
\noindent
We consider the electro-weak sector of the standard model up to the second order
of the perturbation theory (in the causal approach) and derive the most general form of the interaction
Lagrangian for an arbitrary number of Higgs fields. The analysis is done in a purely quantum
setting. If more than one Higgs field is considered, the values of the Weinberg is not fixed uniquely.

\end{abstract}
\end{titlepage}

\section{Introduction}

The general framework of perturbation theory consists in the construction of 
the chronological products such that Bogoliubov axioms are verified \cite{BS},
\cite{EG}, \cite{DF}; for every set of Wick monomials 
$ 
W_{1}(x_{1}),\dots,W_{n}(x_{n}) 
$
acting in the Fock space
$
{\cal H}
$
one associates the distribution-valued operators
$ 
T^{W_{1},\dots,W_{n}}(x_{1},\dots,x_{n}) 
\equiv
T(W_{1}(x_{1}),\dots,W_{n}(x_{n})) 
$ 
called chronological products. The construction of the chronological products
can be done recursively according to Epstein-Glaser prescription \cite{EG},
\cite{Gl} (which reduces the induction procedure to a distribution splitting of 
some distributions with causal support) or according to Stora prescription
\cite{Sto1} (which reduces the renormalization procedure to the process of
extension of
distributions). These products are not uniquely defined but there are some
natural 
limitation on the arbitrariness. If the arbitrariness does not grow with $n$ we
have a
renormalizable theory. An equivalent point of view uses retarded products
\cite{St1}.

Gauge theories describe particles of higher spin. Usually such theories are not
renormalizable. However, one can save renormalizablility using ghost fields.
Such theories are defined in a Fock space
$
{\cal H}
$
with indefinite metric, generated by physical and un-physical fields (called
{\it ghost fields}). One selects the physical states assuming the existence of
an operator $Q$ called {\it gauge charge} which verifies
$
Q^{2} = 0
$
and such that the {\it physical Hilbert space} is by definition
$
{\cal H}_{\rm phys} \equiv Ker(Q)/Im(Q).
$
The space
$
{\cal H}
$
is endowed with a grading (usually called {\it ghost number}) and by
construction the gauge charge is raising the ghost number of a state. Moreover,
the space of Wick monomials in
$
{\cal H}
$
is also endowed with a grading which follows by assigning a ghost number to
every one of the free fields generating
$
{\cal H}.
$
The graded commutator
$
d_{Q}
$
of the gauge charge with any operator $A$ of fixed ghost number
\be
d_{Q}A = [Q,A]
\ee
is raising the ghost number by a unit. It means that
$
d_{Q}
$
is a co-chain operator in the space of Wick polynomials. From now on
$
[\cdot,\cdot]
$
denotes the graded commutator.
 
A gauge theory assumes also that there exists a Wick polynomial of null ghost
number
$
T(x)
$
called {\it the interaction Lagrangian} such that
\be
~[Q, T] = i \partial_{\mu}T^{\mu}
\label{gau1}
\ee
for some other Wick polynomials
$
T^{\mu}.
$
This relation means that the expression $T$ leaves invariant the physical
states, at least in the adiabatic limit. Indeed, if this is true we have:
\be
T(f)~{\cal H}_{\rm phys}~\subset~~{\cal H}_{\rm phys}  
\label{gau2}
\ee
up to terms which can be made as small as desired (making the test function $f$
flatter and flatter). In all known models one finds out that there exist a chain
of Wick polynomials
$
T^{\mu},~T^{\mu\nu},~T^{\mu\nu\rho},\dots
$
such that:
\be
~[Q, T] = i \partial_{\mu}T^{\mu}, \quad
[Q, T^{\mu}] = i \partial_{\nu}T^{\mu\nu}, \quad
[Q, T^{\mu\nu}] = i \partial_{\rho}T^{\mu\nu\rho},\dots
\label{descent}
\ee
It so happens that for all these models the expressions
$
T^{\mu\nu},~T^{\mu\nu\rho},\dots
$
are completely antisymmetric in all indexes; it follows that the chain of
relation stops at the step $4$ (if we work in four dimensions). We can also use
a compact notation
$
T^{I}
$
where $I$ is a collection of indexes
$
I = [\nu_{1},\dots,\nu_{p}]~(p = 0,1,\dots,)
$
and the brackets emphasize the complete antisymmetry in these indexes. All these
polynomials have the same canonical dimension
\be
\omega(T^{I}) = \omega_{0},~\forall I
\ee
and because the ghost number of
$
T \equiv T^{\emptyset}
$
is supposed null, then we also have:
\be
gh(T^{I}) = |I|.
\ee
One can write compactly the relations (\ref{descent}) as follows:
\be
d_{Q}T^{I} = i~\partial_{\mu}T^{I\mu}.
\label{descent1}
\ee

For concrete models the equations (\ref{descent}) can stop earlier: for 
instance in the Yang-Mills case we have
$
T^{\mu\nu\rho} = 0
$
and in the case of gravity
$
T^{\mu\nu\rho\sigma} = 0.
$

Now we can construct the chronological products
$$
T^{I_{1},\dots,I_{n}}(x_{1},\dots,x_{n}) \equiv
T(T^{I_{1}}(x_{1}),\dots,T^{I_{n}}(x_{n}))
$$
according to the recursive procedure. We say that the theory is gauge invariant
in all orders of the perturbation theory if the following set of identities
generalizing (\ref{descent1}):
\be
d_{Q}T^{I_{1},\dots,I_{n}} = 
i \sum_{l=1}^{n} (-1)^{s_{l}} \frac{\partial}{\partial x^{\mu}_{l}}
T^{I_{1},\dots,I_{l}\mu,\dots,I_{n}}
\label{gauge}
\ee
are true for all 
$n \in \N$
and all
$
I_{1}, \dots, I_{n}.
$
Here we have defined
\be
s_{l} \equiv \sum_{j=1}^{l-1} |I|_{j}.
\ee
In particular, the case
$
I_{1} = \dots = I_{n} = \emptyset
$
it is sufficient for the gauge invariance of the scattering matrix, at least
in the adiabatic limit: we have the same argument as for relation (\ref{gau2}).

Such identities can be usually broken by {\it anomalies} i.e. expressions of the
type
$
A^{I_{1},\dots,I_{n}}
$
which are quasi-local and might appear in the right-hand side of the relation
(\ref{gauge}). If one eliminates the anomalies, some restrictions must be
imposed on the interaction Lagrangean, besides those following from
(\ref{gau1}).

In this paper we consider all these restrictions up to the second order of the
perturbation theory and determine the most general form for $T$. This problem
was previously analyzed in great detail in \cite{Sc2}, but no general solution
was found. We find three type of solutions corresponding to the parameter
$
\gamma \equiv \frac{m_{3}~\cos\theta}{m_{1}}
$
taking the values $1$, $> 1$ and $< 1$ respectively. The first case is 
relevant for the usual standard model; more precisely we are investigating 
here only the electro-weak sector. To discriminate between these possibilities 
we can use the super-renormalizablility assumption introduced in \cite{sr3}. In this
way, only the case
$
\gamma = 1
$
remains, but we have to add some extra Higgs scalars.

In the next Section we provide the basis of the causal formalism and its use 
for the standard model. In Section \ref{sm} we describe in detail the conditions
obtained from gauge invariance in the first two orders of perturbation theory.
In Section \ref{sol1} we obtain three types of basic solutions. In Section \ref{symplectic}
we derive the real, antisymmetric, irreducible representations of the group
$
so(3)
$
which are needed for the general solution.
\newpage
\section{The Cohomology of the Gauge Charge Operator\label{q}}

\subsection{Massless Particles of Spin $1$ (Photons)}

We consider a vector space 
$
{\cal H}
$
of Fock type generated (in the sense of Borchers theorem) by the vector field 
$
v_{\mu}
$ 
(with Bose statistics) and the scalar fields 
$
u, \tilde{u}
$
(with Fermi statistics). The Fermi fields are usually called {\it ghost fields}.
We suppose that all these (quantum) fields are of null mass. Let $\Omega$ be the
vacuum state in
$
{\cal H}.
$
In this vector space we can define a sesquilinear form 
$<\cdot,\cdot>$
in the following way: the (non-zero) $2$-point functions are by definition:
\bea
<\Omega, v_{\mu}(x_{1}) v_{\mu}(x_{2})\Omega> =i~\eta_{\mu\nu}~D_{0}^{(+)}(x_{1}
- x_{2}),
\nonumber \\
<\Omega, u(x_{1}) \tilde{u}(x_{2})\Omega> =- i~D_{0}^{(+)}(x_{1} - x_{2})
\qquad
<\Omega, \tilde{u}(x_{1}) u(x_{2})\Omega> = i~D_{0}^{(+)}(x_{1} - x_{2})
\eea
and the $n$-point functions are generated according to Wick theorem. Here
$
\eta_{\mu\nu}
$
is the Minkowski metrics (with diagonal $1, -1, -1, -1$) and 
$
D_{0}^{(+)}
$
is the positive frequency part of the Pauli-Jordan distribution
$
D_{0}
$
of null mass. To extend the sesquilinear form to
$
{\cal H}
$
we define the conjugation by
\be
v_{\mu}^{\dagger} = v_{\mu}, \qquad 
u^{\dagger} = u, \qquad
\tilde{u}^{\dagger} = - \tilde{u}.
\ee

Now we can define in 
$
{\cal H}
$
the operator $Q$ according to the following formulas:
\bea
~[Q, v_{\mu}] = i~\partial_{\mu}u,\qquad
[Q, u] = 0,\qquad
[Q, \tilde{u}] = - i~\partial_{\mu}v^{\mu}
\nonumber \\
Q\Omega = 0
\label{Q-0}
\eea
where by 
$
[\cdot,\cdot]
$
we mean the graded commutator. One can prove that $Q$ is well defined. Indeed,
we have the causal commutation relations 
\be
~[v_{\mu}(x_{1}), v_{\mu}(x_{2}) ] =i~\eta_{\mu\nu}~D_{0}(x_{1} - x_{2})~\cdot
I,
\qquad
[u(x_{1}), \tilde{u}(x_{2})] = - i~D_{0}(x_{1} - x_{2})~\cdot I
\ee
and the other commutators are null. The operator $Q$ should leave invariant
these relations, in particular 
\be
[Q, [ v_{\mu}(x_{1}),\tilde{u}(x_{2})]] + {\rm cyclic~permutations} = 0
\ee
which is true according to (\ref{Q-0}). It is useful to introduce a grading in 
$
{\cal H}
$
as follows: every state which is generated by an even (odd) number of ghost
fields and an arbitrary number of vector fields is even (resp. odd). We denote
by 
$
|f|
$
the ghost number of the state $f$. We notice that the operator $Q$ raises the
ghost number of a state (of fixed ghost number) by an unit. The usefulness of
this construction follows from:
\begin{thm}
The operator $Q$ verifies
$
Q^{2} = 0.
$ 
The factor space
$
Ker(Q)/Ran(Q)
$
is isomorphic to the Fock space of particles of zero mass and helicity $1$
(photons). 
\end{thm}

\newpage

\subsection{Massive Particles of Spin $1$ (Heavy Bosons)}

We repeat the whole argument for the case of massive photons i.e. particles of
spin $1$ and positive mass. 

We consider a vector space 
$
{\cal H}
$
of Fock type generated (in the sense of Borchers theorem) by the vector field 
$
v_{\mu},
$ 
the scalar field 
$
\Phi
$
(with Bose statistics) and the scalar fields 
$
u, \tilde{u}
$
(with Fermi statistics). We suppose that all these (quantum) fields are of mass
$
m > 0.
$
In this vector space we can define a sesquilinear form 
$<\cdot,\cdot>$
in the following way: the (non-zero) $2$-point functions are by definition:
\bea
<\Omega, v_{\mu}(x_{1}) v_{\mu}(x_{2})\Omega> =i~\eta_{\mu\nu}~D_{m}^{(+)}(x_{1}
- x_{2}),
\quad
<\Omega, \Phi(x_{1}) \Phi(x_{2})\Omega> =- i~D_{m}^{(+)}(x_{1} - x_{2})
\nonumber \\
<\Omega, u(x_{1}) \tilde{u}(x_{2})\Omega> =- i~D_{m}^{(+)}(x_{1} - x_{2}),
\qquad
<\Omega, \tilde{u}(x_{1}) u(x_{2})\Omega> = i~D_{m}^{(+)}(x_{1} - x_{2})
\eea
and the $n$-point functions are generated according to Wick theorem. Here
$
D_{m}^{(+)}
$
is the positive frequency part of the Pauli-Jordan distribution
$
D_{m}
$
of mass $m$. To extend the sesquilinear form to
$
{\cal H}
$
we define the conjugation by
\be
v_{\mu}^{\dagger} = v_{\mu}, \qquad 
u^{\dagger} = u, \qquad
\tilde{u}^{\dagger} = - \tilde{u},
\qquad \Phi^{\dagger} = \Phi.
\ee

Now we can define in 
$
{\cal H}
$
the operator $Q$ according to the following formulas:
\bea
~[Q, v_{\mu}] = i~\partial_{\mu}u,\qquad
[Q, u] = 0,\qquad
[Q, \tilde{u}] = - i~(\partial_{\mu}v^{\mu} + m~\Phi)
\qquad
[Q,\Phi] = i~m~u,
\nonumber \\
Q\Omega = 0.
\label{Q-m}
\eea
One can prove that $Q$ is well defined. We have a result similar to the first
theorem of this Section:
\begin{thm}
The operator $Q$ verifies
$
Q^{2} = 0.
$ 
The factor space
$
Ker(Q)/Ran(Q)
$
is isomorphic to the Fock space of particles of mass $m$ and spin $1$ (massive
photons). 
\end{thm}
\newpage
\subsection{The Generic Yang-Mills Case}

The situations described above (of massless and massive photons) are susceptible
of the following generalizations. We can consider a system of 
$
r_{1}
$ 
species of particles of null mass and helicity $1$ if we use in the first part
of this Section 
$
r_{1}
$ 
triplets
$
(v^{\mu}_{a}, u_{a}, \tilde{u}_{a}), a \in I_{1}
$
of massless fields; here
$
I_{1}
$
is a set of indexes of cardinal 
$
r_{1}.
$
All the relations have to be modified by appending an index $a$ to all these
fields. 

In the massive case we have to consider 
$
r_{2}
$ 
quadruples
$
(v^{\mu}_{a}, u_{a}, \tilde{u}_{a}, \Phi_{a}),  a \in I_{2}
$
of fields of mass 
$
m_{a}
$; here
$
I_{2}
$
is a set of indexes of cardinal 
$
r_{2}.
$

We can consider now the most general case involving fields of spin not greater
that $1$.
We take 
$
I = I_{1} \cup I_{2} \cup I_{3}
$
a set of indexes and for any index we take a quadruple
$
(v^{\mu}_{a}, u_{a}, \tilde{u}_{a},\Phi_{a}), a \in I
$
of fields with the following conventions:
(a) For
$
a \in I_{1}
$
we impose 
$
\Phi_{a} = 0
$
and we take the masses to be null
$
m_{a} = 0;
$
(b) For
$
a \in I_{2}
$
we take the all the masses strictly positive:
$
m_{a} > 0;
$
(c) For 
$
a \in I_{3}
$
we take 
$
v_{a}^{\mu}, u_{a}, \tilde{u}_{a}
$
to be null and the fields
$
\Phi_{a} \equiv \phi^{H}_{a} 
$
of mass 
$
m_{a} \equiv m^{H}_{a} \geq 0.
$
The fields
$
\phi^{H}_{a} 
$
are called {\it Higgs fields}.

If we define
$
m_{a} = 0, \forall a \in I_{3}
$
then we can define in 
$
{\cal H}
$
the operator $Q$ according to the following formulas for all indexes
$
a \in I:
$
\bea
~[Q, v^{\mu}_{a}] = i~\partial^{\mu}u_{a},\qquad
[Q, u_{a}] = 0,
\nonumber \\
~[Q, \tilde{u}_{a}] = - i~(\partial_{\mu}v^{\mu}_{a} + m_{a}~\Phi_{a})
\qquad
[Q,\Phi_{a}] = i~m_{a}~u_{a},
\nonumber \\
Q\Omega = 0.
\label{Q-general}
\eea

If we consider matter fields also i.e some set of Dirac fields with Fermi
statistics:
$
\Psi_{A}, A \in I_{4}
$ 
then we impose
\be
d_{Q}\Psi_{A} = 0 
\ee
and the space 
$
{\cal P}_{0}
$
is generated by
$
\Psi_{A}
$
and
$
\bar{\Psi}_{A}
$
also.

\subsection{The Yang-Mills Lagrangian}

Now we consider the framework and notations from the end of the preceding
Section. Then we have the following
result which describes the most general form of the Yang-Mills interaction.
Summation over the dummy indexes is used everywhere.

Let $T$ be a relative cocycle for 
$
d_{Q}
$
which is tri-linear in the fields and is of canonical dimension
$
\omega(T) \leq 4
$
and ghost number
$
gh(T) = 0.
$
Then:
(i) $T$ is (relatively) cohomologous to a non-trivial co-cycle of the form:
\bea
T = f_{abc} \left( \frac{1}{2}~v_{a\mu}~v_{b\nu}~F_{c}^{\nu\mu}
+ u_{a}~v_{b}^{\mu}~d_{\mu}\tilde{u}_{c}\right)
\nonumber \\
+ f^{\prime}_{abc} (\Phi_{a}~\phi_{b}^{\mu}~v_{c\mu} +
m_{b}~\Phi_{a}~\tilde{u}_{b}~u_{c})
\nonumber \\
+ \frac{1}{3!}~f^{\prime\prime}_{abc}~\Phi_{a}~\Phi_{b}~\Phi_{c}
+ j^{\mu}_{a}~v_{a\mu} + j_{a}~\Phi_{a};
\label{T-sm}
\eea

(ii) The relation 
$
d_{Q}T = i~\partial_{\mu}T^{\mu}
$
is verified by:
\be
T^{\mu} = f_{abc} \left( u_{a}~v_{b\nu}~F^{\nu\mu}_{c} -
\frac{1}{2} u_{a}~u_{b}~d^{\mu}\tilde{u}_{c} \right)
+ f^{\prime}_{abc}~\Phi_{a}~\phi_{b}^{\mu}~u_{c}
+ j^{\mu}_{a}~u_{a}
\label{Tmu}
\ee

(iii) The relation 
$
d_{Q}T^{\mu} = i~\partial_{\nu}T^{\mu\nu}
$
is verified by:
\be
T^{\mu\nu} \equiv \frac{1}{2} f_{abc}~u_{a}~u_{b}~F_{c}^{\mu\nu}.
\ee

Here
\be
\phi_{a}^{\mu} \equiv \partial^{\mu}\phi_{a} - m~v_{a}^{\mu}.
\ee

There are various restrictions on the constants appearing in the preceding 
expressions. We are interested in the structure of the coefficients
$
f_{abc}
$
and
$
f^{\prime}_{abc}
$
determining the electro-weak sector. We can imposed the following restrictions:

\be
f^{\prime}_{abc} = - (a \leftrightarrow b)
\label{c1}
\ee
\be
f^{\prime}_{abc} = 0,\qquad \forall c \in I_{3}
\label{c2}
\ee
\be
f^{\prime}_{abc} = 0,\qquad \forall a \in I_{1}
\label{c3}
\ee

The preceding expressions
$
T^{I}
$
are self-adjoint if the constants
$
f_{abc},~f^{\prime}_{abc}
$
are real.

\subsection{First and Second Order Gauge Invariance\label{second}}

It can be proved that first order gauge invariance (in the electro-weak sector) leads to:
\be
f^{\prime}_{cab}~m_{a} - f^{\prime}_{cba}~m_{b}
= f_{abc}~m_{c},~\forall a,b \in I_{1} \cup I_{2},~c \in I_{2} \cup I_{3}
\label{f1}
\ee
 From here:
\be
f^{\prime}_{abc} = \frac{m_{a}^{2} + m_{b}^{2} - m_{c}^{2}}{2 m_{a}
m_{b}}~f_{abc}, 
\qquad a, b, c \in I_{2}
\label{f11}
\ee
\be
f_{abc}~(m_{a} - m_{b}) = 0,\qquad c \in I_{1},~a,b \in I_{2}
\label{f12}
\ee
\be
f^{\prime}_{abc} = f_{abc}, \qquad a, b \in I_{2}, c \in I_{1}
\label{f13}
\ee
\be
f^{\prime}_{abc} = m_{c}~g_{abc},~g_{abc} = g_{acb},
\qquad a \in I_{3},~b,c \in I_{2}
\label{f14}
\ee
\be
f^{\prime}_{abc} = 0,
\qquad
a \in I_{3},~b \in I_{2}, c \in I_{1}.
\label{f15}
\ee

It also can be proved that second order gauge invariance (in the electro-weak sector) leads to:

\be
\sum_{c}~(f_{abc}~f_{dec} + f_{bdc}~f_{aec} + f_{dac}~f_{bec}) = 0
\label{jacoby}
\ee
(which is the Jacobi identity) and
\be
\sum_{c}~[ f^{\prime}_{dca}~f^{\prime}_{ceb}
- (a \leftrightarrow b) ] = 
- \sum_{c}~f_{abc}~f^{\prime}_{dec},
\qquad
a,b \in I_{1} \cup I_{2},~d,e \in I_{2} \cup I_{3}.
\label{f21}
\ee

\subsection{Explicit Form of (\ref{f21})}

To simplify the analysis of the restrictions listed above we use the vectorial notations and we denote the indices from
$I_{3}$
by
$j,k,...$
We define the vectors
$
g_{ab} \in \R^{|I_{3}|}
$
by:
\be
(g_{ab})_{j} \equiv g_{jab},\quad \forall j \in I_{3},~\forall a,b = 1,2,3
\ee
and according to (\ref{f14}) we have symmetry in 
$
a \leftrightarrow b.
$
We also define the matrices
$
T_{a} \in M_{\R}(|I_{3}|,|I_{3}|)
$
according to:
\be
(T_{a})_{jk} \equiv f^{\prime}_{jka}
\ee
and observe that these matrices are antisymmetric. Finally we define
$
G \in M_{\R}(|I_{2}|,|I_{2}|)
$
by
\be
G_{ab} \equiv m_{a}~m_{b}~\sum_{c \in I_{2}}~(g_{ac}~g_{bc}^{T} -
g_{bc}~g_{ac}^{T}),\quad \forall a,b \in I_{2}
\ee
and note that we have symmetry in 
$
a \leftrightarrow b.
$
From (\ref{f21}) we obtain in detail the following cases:

1) $a, b, d, e \in I_{2}$
\bea
m_{a}~m_{b}~(<g_{ad},g_{be}> - <g_{bd},g_{ae}> ) =
\nonumber \\
\left[ \sum_{c \in I_{2}} f^{\prime}_{dca}~f^{\prime}_{ceb} - (a \leftrightarrow
b) \right] 
+ \sum_{c \in I_{1}} f_{abc}~f_{dec} + \sum_{c \in I_{2}}
f_{abc}~f^{\prime}_{dec}
\label{f21-1}
\eea

2) $a, b, d \in I_{2},~ e \in I_{3}$

\be
m_{a}~T_{b}~g_{ad} - m_{b}~T_{a}~g_{bd} =
\nonumber\\
\sum_{c \in I_{2}}~(m_{c}~f_{abc}~g_{cd} - m_{a}~f^{\prime}_{dcb}~g_{ac} +
m_{b}~f^{\prime}_{dca}~g_{bc}) 
\label{f21-2}
\ee

3) $a, b \in I_{2},~ d, e \in I_{3}$

\be
T_{a}~T_{b} - T_{b}~T_{a} - G_{ab} = -
\sum_{c \in I_{1} \cup I_{2}}~f_{abc}~T_{c}
\label{f21-3}
\ee

4) $a \in I_{1},~ b, d \in I_{2}, e \in I_{3}$

\be
T_{a}~g_{bd} = - \sum_{c \in I_{2}} (f_{abc}~g_{cd} + f_{adc}~g_{bc})
\label{f21-4}
\ee

5) $a \in I_{1}, b \in I_{2},~ d, e \in I_{3}$

\be
T_{a}~T_{b} - T_{b}~T_{a} = - \sum_{c \in I_{1} \cup I_{2}}~f_{abc}~T_{c}
\label{f21-5}
\ee

6) $a \in I_{1}, b, d, e \in I_{3}$

\be
\sum_{c \in I_{1}}~f_{abc}~f_{dec} = 0
\label{f21-6}
\ee

This is the only non-trivial case requiring some computations.

7) $a, b \in I_{1}, d, e \in I_{2}$

\be
\left[ \sum_{c \in I_{2}} f_{dca}~f_{ceb} - ( a \leftrightarrow b) \right] 
+ \sum_{c \in I_{1}} f_{abc}~f_{dec} + \sum_{c \in I_{2}}
f_{abc}~f^{\prime}_{dec} = 0
\label{f21-7}
\ee

8) $a, b \in I_{1}, d \in I_{2}, e \in I_{3}$

\be
\sum_{c \in I_{2}} f_{abc}~g_{cd} = 0
\label{f21-8}
\ee

9) $a, b \in I_{1}, d, e \in I_{3}$

\be
T_{a}~T_{b} - T_{b}~T_{a} = - \sum_{c \in I_{1} \cup I_{2}}~f_{abc}~T_{c}
\label{f21-9}
\ee
\newpage

\section{The Standard Model\label{sm}}

We consider the following particular case relevant for the electro-weak
sector of the standard model. The Lie algebra is real and isomorphic to
$
u(1) \times su(2)
$
and we have
$
I_{1} = \{0\}, I_{2} = \{1,2,3\}.
$
The non-zero constants
$
f_{abc}
$
are:
\be
f_{210} = \sin\theta, \quad f_{321} = \cos\theta
\ee
with
$\cos \theta > 0$
and the other constants determined through the anti-symmetry property;
$
\theta
$
is the {\it Weinberg angle}. It is interesting to see that for a
four-dimensional
Lie algebra, the Jacobi identity is trivially verified. So there are two cases:
only one of the structure constants
$
f_{012}, f_{023}, f_{031}
$
is non-zero (and we end up with the case above after some re-scalings) and the 
case when at least two of the preceding structure constants are non-zero. The
last case leads to the equality of all masses and it is not interesting from the
physical point of view.

1) We consider first order gauge invariance:

- From (\ref{f12}) we obtain
\be
m_{1} = m_{2}
\label{m1}
\ee

- From the other identities we get:
\bea
f^{\prime}_{231} = f^{\prime}_{312} = - \cos\theta~\frac{m_{3}}{2 m_{1}}
\nonumber \\
f^{\prime}_{123} = - \cos\theta~\left( 1 - \frac{m_{3}^{2}}{2 m_{1}^{2}} \right)
\nonumber\\
f^{\prime}_{120} = - \sin\theta
\label{f12-1}
\eea
\bea
f^{\prime}_{121} = f^{\prime}_{131} = f^{\prime}_{122} = f^{\prime}_{232} =
f^{\prime}_{133} =
f^{\prime}_{233} = f^{\prime}_{130} = f^{\prime}_{230} = 0
\nonumber \\
f^{\prime}_{0bc} = 0, \forall b, c 
\nonumber \\
f^{\prime}_{j10} = f^{\prime}_{j20} = f^{\prime}_{j30} = 0,~\forall j \in I_{3} 
\label{f12-2}
\eea
\bea
f^{\prime}_{jab} = m_{a}~g_{jab},
\nonumber\\
g_{jab} = g_{jba}, \qquad \forall j \in I_{3}, \forall a, b = 1,2,3.
\label{f12-3}
\eea

2) Second order gauge invariance. We rewrite the relations 1)-9) from the 
preceding Section defining by
$
< \cdot, \cdot >
$
and
$
| \cdot |
$
the scalar product and respectively the norm from
$
\R^{|I_{3}|}.
$
Then we have from (\ref{f21}) the following cases:

1) $a, b, d, e \in I_{2}$

\be
m_{1}^{2}(<g_{11},g_{22}> - |g_{12}|^{2}) = 1 - \cos^{2}\theta~
\frac{3 m_{3}^{2}}{4 m_{1}^{2}}
\label{f21-1.1}
\ee
\be
m_{1}^{2}(<g_{22},g_{33}> - |g_{23}|^{2}) = \cos^{2}\theta~
\frac{m_{3}^{2}}{4 m_{1}^{2}}
\label{f21-1.2}
\ee
\be
m_{1}^{2}(<g_{11},g_{33}> - |g_{13}|^{2}) = \cos^{2}\theta~
\frac{m_{3}^{2}}{4 m_{1}^{2}}
\label{f21-1.3}
\ee
\be
<g_{11},g_{23}> - <g_{12},g_{13}> = 0
\label{f21-1.4}
\ee
\be
<g_{22},g_{13}> - <g_{12},g_{23}> = 0
\label{f21-1.5}
\ee
\be
<g_{12},g_{33}> - <g_{13},g_{23}> = 0
\label{f21-1.6}
\ee

2) $a, b, d\in I_{2},~~e \in I_{3}$

\be
m_{1}~(T_{2}~g_{11} - T_{1}~g_{12}) = 
- \frac{3}{2}~m_{3}~\cos\theta~g_{13}
\label{f21-2.1}
\ee
\be
m_{1}~(T_{2}~g_{12} - T_{1}~g_{22}) = 
- \frac{3}{2}~m_{3}~\cos\theta~g_{23}
\label{f21-2.2}
\ee
\be
m_{1}~(T_{2}~g_{13} - T_{1}~g_{23}) = 
m_{3}~\cos\theta~\left(\frac{1}{2}~g_{11} + \frac{1}{2}~g_{22} - g_{33} \right)
\label{f21-2.3}
\ee
\be
T_{3}~g_{12} - \frac{m_{3}}{m_{1}}~T_{2}~g_{13} = 
\cos\theta~\left[ - g_{11} + \left( 1 - \frac{m_{3}^{2}}{2 m_{1}^{2}}
\right)~g_{22} + \frac{m_{3}^{2}}{2 m_{1}^{2}} g_{33} \right]
\label{f21-2.4}
\ee
\be
T_{3}~g_{22} - \frac{m_{3}}{m_{1}}~T_{2}~g_{23} = 
- \cos\theta~\left( 2 - \frac{m_{3}^{2}}{2 m_{1}^{2}} \right)~g_{12}
\label{f21-2.5}
\ee
\be
T_{3}~g_{23} - \frac{m_{3}}{m_{1}}~T_{2}~g_{33} = 
- \cos\theta~\left( 1 + \frac{m_{3}^{2}}{2 m_{1}^{2}} \right)~g_{13}
\label{f21-2.6}
\ee
\be
T_{3}~g_{11} - \frac{m_{3}}{m_{1}}~T_{1}~g_{13} = 
\cos\theta~\left( 2 - \frac{m_{3}^{2}}{2 m_{1}^{2}} \right)~g_{12}
\label{f21-2.7}
\ee
\be
T_{3}~g_{13} - \frac{m_{3}}{m_{1}}~T_{1}~g_{33} = 
\cos\theta~\left( 1 + \frac{m_{3}^{2}}{2 m_{1}^{2}} \right)~g_{23}
\label{f21-2.8}
\ee
\be
T_{3}~g_{12} - \frac{m_{3}}{m_{1}}~T_{1}~g_{23} = 
\cos\theta~\left[ - \left( 1 - \frac{m_{3}^{2}}{2 m_{1}^{2}} \right)~g_{11} +
g_{22} - \frac{m_{3}^{2}}{2 m_{1}^{2}} g_{33} \right]
\label{f21-2.9}
\ee

3) $a, b \in I_{2},~d,e \in I_{3}$
\be
T_{1}~T_{2} - T_{2}~T_{1} - G_{12} = \sin\theta~T_{0} + \cos\theta~T_{3}
\label{f21-3.1}
\ee
\be
T_{2}~T_{3} - T_{3}~T_{2} - G_{23} = \cos\theta~T_{1}
\label{f21-3.2}
\ee
\be
T_{3}~T_{1} - T_{1}~T_{3} - G_{23} = \cos\theta~T_{2}
\label{f21-3.3}
\ee

4) $a \in I_{1},~b,d \in I_{2},~e \in I_{3}$

\be
T_{0}~g_{11} = 2 \sin\theta~g_{12}
\label{f21-4.1}
\ee
\be
T_{0}~g_{12} = \sin\theta~( - g_{11} + g_{22})
\label{f21-4.2}
\ee
\be
T_{0}~g_{22} = - 2 \sin\theta~g_{12}
\label{f21-4.3}
\ee
\be
T_{0}~g_{13} = \sin\theta~g_{23}
\label{f21-4.4}
\ee
\be
T_{0}~g_{23} = - \sin\theta~g_{13}
\label{f21-4.5}
\ee
\be
T_{0}~g_{33} = 0.
\label{f21-4.6}
\ee

5) $a \in I_{1},~b \in I_{2},~d,e \in I_{3}$

\be
T_{0}~T_{1} - T_{1}~T_{0} = \sin\theta~T_{2}
\label{f21-5.1}
\ee
\be
T_{0}~T_{2} - T_{2}~T_{1} = - \sin\theta~T_{1}
\label{f21-5.2}
\ee
\be
T_{0}~T_{3} - T_{3}~T_{0} = 0
\label{f21-5.3}
\ee

6) $a \in I_{1},~b,d,e \in I_{2}$

We obtain identities.

Not all of the preceding identities are independent. From the relations
(\ref{f21-3.1}) - (\ref{f21-3.3}) and (\ref{f21-5.1}) - (\ref{f21-5.3})
we see that the expressions
$
T_{a}
$
are a representation, up to the cocycle
$
G_{ab}.
$

\newpage
\section{Solution of the System\label{sol1}}

In the space
$
\R^{|I_{3}|}
$
we identify the subspace $V$ generated by applying polynomials in
$
I, T_{0}, \dots, T_{3}
$
on the vectors 
$
g_{ab}.
$
It is obvious that this space is left invariant by the operators
$
T_{0}, \dots, T_{3}.
$
Moreover because of the antisymmetry property:
\be
<x, T_{a}y> = - <T_{a}x, y>, \quad \forall x, y \in \R^{|I_{3}|}
\label{symF}
\ee
of the matrices
$
T_{a}
$
it follows that the orthogonal subspace 
$
V^{\bot}
$
(with respect to the scalar product
$
< \cdot, \cdot >)
$
is also left invariant by the operators
$
T_{0}, \dots, T_{3}
$
and the restriction of these operators to 
$
V^{\bot}
$
is a representation of the algebra
$
su(1) \times su(2).
$

So we must divide the analysis in two parts corresponding to the two subspaces $V$ and 
$
V^{\bot}.
$
We immediately prove that we cannot have
$
V = \{ 0 \}
$
so we start the analysis in the subspace $V$. First we have
\begin{thm}
The system of vectors
$
g_{\pm} \equiv \frac{1}{2}~(g_{11} \pm g_{22)},~g_{12},~g_{13},~g_{23}
$
is ortho-normal. 
\end{thm}
{\bf Proof:}
We can rewrite the system (\ref{f21-4.1}) - (\ref{f21-4.6}) as:
\be
T_{0}~g_{-} = 2 \sin\theta~g_{12}
\label{0-1}
\ee
\be
T_{0}~g_{12} = - 2 \sin\theta~g_{-}
\label{0-2}
\ee
\be
T_{0}~g_{13} = \sin\theta~g_{23}
\label{0-3}
\ee
\be
T_{0}~g_{23} = - \sin\theta~g_{13}
\label{0-4}
\ee
\be
T_{0}~g_{+} = 0.
\label{0-5}
\ee

We use the antisymmetry  property (\ref{symF}) of the matrix
$
T_{0}
$.
It is known that for such a matrix we can find a basis
$
E_{0}^{j}
$
and
$
E_{A}^{k}, F_{A}^{k}
$
such that
\be
T_{0}~E_{0}^{j} = 0
\ee
and
\be
T_{0}E^{k}_{A} = - \lambda_{A}~F^{k}_{A}, \qquad
T_{0}F^{k}_{A} = \lambda_{A}~F^{k}_{A}
\label{EF-generic}
\ee
where
$
\lambda_{A} > 0
$
are distinct positive eigenvalues of the operator
$
- (T_{0})^{1/2}
$.
It is now easy to prove that the basis above is an orthonormal system.
Moreover we note that:
\be
|E^{j}_{A}| = |F^{j}_{A}|.
\ee
Now we see that the vector
$
g_{+}
$
is associated with the eigenvalue $0$, the vectors
$
g_{13}, g_{23}
$
with the eigenvalue $1$ and the vectors
$
g_{-}, g_{12}
$
with the eigenvalue $2$ and this leads to the result.
$\qed$

The eigenvalue subspaces of
$
T_{0}
$
are more precisely described as follows.
\begin{thm}
Let us define
\be
A \equiv - (T_{0})^{2}.
\ee

Then:

(i) The eigenvalues of $A$ are $\geq 0$.

(ii) If
$
E \in V_{\lambda^{2}}
$
then
$$
T_{1}E, T_{2}E \in V_{(\lambda + \sin\theta)^{2}} \oplus
V_{(\lambda - \sin\theta)^{2}}.
$$
Moreover
$$
T_{3} V_{\lambda^{2}} \subset V_{\lambda^{2}}.
$$
(iii) The subspace $V$ can be written as
\be
V = \oplus_{m \geq 0}~V_{m}
\ee
where
\be
A|_{V_{m}} = m^{2}~\sin^{2}\theta \cdot I
\ee
\label{f012}
\end{thm}
{\bf Proof:} Because of the antisymmetry of the matrix
$
T_{0}
$
we consider, as above, a basis 
$
E^{j}_{\lambda}, F^{j}_{\lambda} \in V_{\lambda^{2}}
$ 
such that
\be
T_{0}E^{j}_{\lambda} = - \lambda~F^{j}_{\lambda}, \qquad
T_{0}F^{j}_{\lambda} = \lambda~F^{j}_{\lambda}
\label{EF}
\ee
for the values
$
\lambda > 0.
$

From (\ref{f21-5.1}) we obtain
\be
T_{2} = \frac{1}{\sin\theta}~(T_{0}~T_{1} - T_{1}~T_{0})
\ee
and if we substitute in (\ref{f21-5.2}) we get:
\be
(T_{0})^{2}~T_{1} + T_{1}~(T_{0})^{2} - 2~T_{0}~T_{1}~T_{0} =
- \sin^{2}\theta~T_{1}
\ee
From here:
\bea
(A + \lambda^{2} - \sin^{2}\theta)~T_{1}~E^{j}_{\lambda} -
2~\lambda~T_{0}~T_{1}~F^{j}_{\lambda} = 0
\nonumber \\
(A + \lambda^{2} - \sin^{2}\theta)~T_{1}~F^{j}_{\lambda} +
2~\lambda_{j}~T_{0}~T_{1}~E^{j}_{\lambda} = 0
\eea

It follows from here:
\bea
[ (A + \lambda^{2} - \sin^{2}\theta)^{2} - 4~\lambda^{2}~A]~T_{1}E^{j}_{\lambda} = 0
\nonumber\\
~[ (A + \lambda^{2} - \sin^{2}\theta)^{2} - 4~\lambda^{2}~A]~T_{1}F^{j}_{\lambda} = 0
\eea
Because the preceding square bracket is
\be
[ \cdots ] = [ A - (\lambda + \sin~\theta)^{2} ]~[ A - (\lambda - \sin~\theta)^{2} ]
\ee
it follows that
$
(\lambda \pm sin~\theta)^{2}
$
can be also eigenvalues of $A$ and this gives (ii) of the theorem.

We also note that
\be
g_{+} \in V_{0},\qquad g_{13},~g_{23} \in V_{1},\qquad g_{-},~g_{12} \in V_{2}.
\label{g-ab}
\ee
So if we use (ii) of the theorem we can generate from the vectors 
$
g_{ab}
$
only vectors from 
$
V_{n}.
$
$\qed$

We need a convenient basis in the real vector space $V$. This basis is obtained in a similar way
to the complex case where one uses raising and lowering operators. However,
working in real vector spaces is more difficult. The basic trick is used for the proof of
the following result:
\begin{thm}
Suppose that
$
V_{0} \not= \{0\}
$
and let
$
\{E_{0}^{j}\}_{j \in J_{0}}
$
be a orthonormat basis (here 
$
J_{0}
$
is some finite index set) i.e.
\be
< E_{0}^{j}, E_{0}^{k} > = \delta_{jk},\quad \forall j,k \in J_{0}
\label{ortho-0}
\ee
and also
\be
T_{0}~E_{0}^{j} = 0,\quad \forall j \in J_{0}
\label{e0}
\ee
Then there exists the chain of subsets of 
$
J_{0}
$
\be
J_{N} \subset J_{N-1} \subset \cdots J_{1} \subset J_{0}
\nonumber
\ee
and the system of orthonormal vectors
$
\{E_{n}^{j}, F_{n}^{j}\}_{j \in J_{n}} 
$
in
$
V_{n}
$
such that
\be
T_{0}~E_{n}^{j} = - n~\sin\theta~F_{n}^{j}, \qquad
T_{0}~F_{n}^{j} = n~\sin\theta~F_{n}^{j},\quad \forall j \in J_{n}.
\label{EFn}
\ee

Let us denote for 
$
n = 0,\dots,N
$
\be
\alpha_{n}^{j} \equiv |E_{n}^{j}|^{2} = |F_{n}^{j}|^{2}, \quad \forall j \in J_{n}
\label{alfa}
\ee
and for 
$
n = 1,\dots,n
$
\be
A_{n}^{j} \equiv \alpha_{n}^{j} / \alpha_{n-1}^{j}, \quad \forall j \in J_{n}. 
\label{An}
\ee
Then we have 

(i) for
$
\forall j \in J_{0}
$
\be
T_{1}~E_{0}^{j} = E_{1}^{j}, \qquad
T_{2}~E_{0}^{j} = - F_{1}^{j},
\label{ef12-0}
\ee
(ii) for
$
\forall j \in J_{1}
$
\bea
T_{1}~E_{1}^{j} = - \alpha_{1}^{j}~E_{0}^{j} + E_{2}^{j}
\nonumber\\
T_{1}~F_{1}^{j} = F_{2}^{j}
\nonumber\\
T_{2}~E_{1}^{j} = - F_{2}^{j}
\nonumber\\
T_{2}~F_{1}^{j} = \alpha_{1}^{j}~E_{0}^{j} + E_{2}^{j}
\label{ef12-1}
\eea
(iii) and for 
$
n = 2,\dots,N
$
and
$
\forall j \in J_{n}
$
\bea
T_{1}~E_{n}^{j} = - A_{n}^{j}~E_{n-1}^{j} + E_{n+1}^{j}
\nonumber\\
T_{1}~F_{n}^{j} = - A_{n}^{j}~F_{n-1}^{j} + F_{n+1}^{j}
\nonumber\\
T_{2}~E_{n}^{j} = - A_{n}^{j}~F_{n-1}^{j} - F_{n+1}^{j}
\nonumber\\
T_{2}~F_{n}^{j} = A_{n}^{j}~E_{n-1}^{j} + E_{n+1}^{j}
\label{ef12-n}
\eea
where we make the convention that 
\bea
E_{n}^{j} = 0,\quad F_{n}^{j} = 0,\quad \forall j \in J_{n-1} - J_{n} 
\nonumber\\
E_{N+1}^{j} = 0,\quad F_{N+1}^{j} = 0
\eea
\label{T12}
\end{thm}
{\bf Proof:} (i) We make the choice of the basis
$
\{E_{0}^{j}\}_{j \in J_{0}}
$
such that we have (\ref{ortho-0}) and (\ref{e0}) and then apply the previous
theorem and get 
$
T_{1}~E_{0}^{j} \in V_{1}
$
so we define
$
E_{1}^{j} \in V_{1}
$
according to
\be
T_{1}~E_{0}^{j} = E_{1}^{j},\quad \forall j \in J_{0}.
\ee
Next, we define 
$
F_{1}^{j} \in V_{1}
$
through
\be
T_{0}~E_{1}^{j} = - \sin\theta~F_{1}^{j},\quad \forall j \in J_{0}
\ee
and by direct computations, using (\ref{f21-5.1}) and (\ref{f21-5.2}) we obtain
\be
T_{2}~E_{0}^{j} = - F_{1}^{j},\quad \forall j \in J_{0}
\ee
and
\be
T_{0}~F_{1}^{j} = \sin\theta~E_{1}^{j},\quad \forall j \in J_{0}.
\ee
Of course, we might have
\be
\alpha_{1}^{j} = 0
\ee
for some values of 
$
j \in J_{0}
$
(and in this case
$
E_{1}^{j} = F_{1}^{j} = 0
$)
so we define
\be
J_{1} = \{j \in J_{0} \quad | \quad \alpha_{1}^{j} \not= 0 \}.
\ee

(ii) Now we apply again the preceding theorem and get that
$
T_{1}~E_{1}^{j}, T_{1}~F_{1}^{j} \in V_{0} \oplus V_{2}.
$
We choose a basis 
$
\{E_{2}^{j}, F_{2}^{j}\}
$
in
$
V_{2}
$
such that we have (\ref{EFn}) for 
$
n = 2
$
and we have 
$
\forall j \in J_{1}
$
the following generic formulas:
\bea
T_{1}~E_{1}^{j} = a_{jk}~E_{0}^{k} + c_{jk}~E_{2}^{k} + d_{jk}~F_{2}^{k} 
\nonumber\\
T_{1}~F_{1}^{j} = \tilde{a}_{jk}~E_{0}^{k} + \tilde{c}_{jk}~E_{2}^{k} + \tilde{d}_{jk}~F_{2}^{k} 
\eea
If we use the antisymmetry of the operator
$
T_{1}
$
we obtain a more precise form:
\bea
T_{1}~E_{1}^{j} = - \alpha^{j}_{1}~E_{0}^{j} + c_{jk}~E_{2}^{k} + d_{jk}~F_{2}^{k} 
\nonumber\\
T_{1}~F_{1}^{j} = \tilde{c}_{jk}~E_{2}^{k} + \tilde{d}_{jk}~F_{2}^{k}.
\eea
Next, we use (\ref{f21-5.1}) to obtain the expression of
$
T_{2}
$:
\bea
T_{2}~E_{1}^{j} = (\tilde{c} + 2 d)_{jk}~E_{2}^{k} + (\tilde{d} - 2c)_{jk}~F_{2}^{k} 
\nonumber\\
T_{1}~F_{1}^{j} = \alpha^{j}_{1}~E_{0}^{j} + (2 \tilde{d} - c)_{jk}~E_{2}^{k} - (d + 2\tilde{c})_{jk}~F_{2}^{k}.
\eea
Now if we use (\ref{f21-5.2}) we obtain the restrictions
\be
\tilde{c} = - d,\quad \tilde{d} = c.
\ee
If we define
\be
\tilde{E}_{2}^{j} \equiv c_{jk}~E_{2}^{k} + d_{jk}~F_{2}^{k},\quad
\tilde{F}_{2}^{j} \equiv = - d_{jk}~E_{2}^{k} + c_{jk}~F_{2}^{k},\quad \forall j \in J_{1}
\ee
we have 
\be
T_{0}~\tilde{E}_{2}^{j} = - 2~\sin\theta~\tilde{F}_{2}^{j}, \qquad
T_{0}~\tilde{F}_{2}^{j} = 2~\sin\theta~E_{2}^{j},\quad \forall j \in J_{1}.
\ee
and
\bea
T_{1}~E_{1}^{j} = - \alpha_{1}^{j}~E_{0}^{j} + \tilde{E}_{2}^{j}
\nonumber\\
T_{1}~F_{1}^{j} = \tilde{F}_{2}^{j}
\nonumber\\
T_{2}~E_{1}^{j} = - \tilde{F}_{2}^{j}
\nonumber\\
T_{2}~F_{1}^{j} = \alpha_{1}^{j}~E_{0}^{j} + \tilde{E}_{2}^{j}
\eea
so if we redefine
\be
\tilde{E}_{2}^{j}\rightarrow E_{2}^{j},\quad \tilde{F}_{2}^{j}\rightarrow F_{2}^{j}
\label{redef}
\ee
we get (\ref{ef12-1}). Of course, we might have
\be
\alpha_{2}^{j} = 0
\ee
for some values of 
$
j \in J_{1}
$
(and in this case
$
E_{2}^{j} = F_{2}^{j} = 0
$)
so we define
\be
J_{2} = \{j \in J_{1} \quad | \quad \alpha_{2}^{j} \not= 0 \}.
\ee
We have obtained the convenient basis in 
$
V_{2}.
$

(iii) Next, we use the preceding theorem and have
$
T_{1}~E_{2}^{j}, T_{1}~F_{2}^{j} \in V_{1} \oplus V_{3}.
$
We choose a basis 
$
\{E_{3}^{j}, F_{3}^{j}\}
$
in
$
V_{3}
$
such that we have (\ref{EFn}) for 
$
n = 3
$
and we have 
$
\forall j \in J_{2}
$
the following generic formulas:
\bea
T_{1}~E_{2}^{j} = a_{jk}~E_{1}^{k} + b_{jk}~F_{1}^{k} + c_{jk}~E_{3}^{k} + d_{jk}~F_{3}^{k} 
\nonumber\\
T_{1}~F_{2}^{j} = \tilde{a}_{jk}~E_{1}^{k} + \tilde{b}_{jk}~F_{1}^{k} + \tilde{c}_{jk}~E_{3}^{k} + \tilde{d}_{jk}~F_{3}^{k}.
\eea
If we use the antisymmetry of the matrix
$
T_{1}
$
we obtain a simplified form:
\bea
T_{1}~E_{2}^{j} = - A_{2}^{j}~E_{1}^{j} + c_{jk}~E_{3}^{k} + d_{jk}~F_{3}^{k} 
\nonumber\\
T_{1}~F_{2}^{j} = - A_{2}^{j}~F_{1}^{j} + \tilde{c}_{jk}~E_{3}^{k} + \tilde{d}_{jk}~F_{3}^{k}.
\eea

As at (ii) we use (\ref{f21-5.1}) to obtain the expression of
$
T_{2}
$:
\bea
T_{2}~E_{2}^{j} = - A_{2}^{j}~F_{1}^{j} + (2 \tilde{c} + 3 d)_{jk}~E_{3}^{k} + (2 \tilde{d} - 3c)_{jk}~F_{2}^{k} 
\nonumber\\
T_{1}~F_{2}^{j} = A_{2}^{j}~E_{1}^{j} + (3 \tilde{d} - 2 c)_{jk}~E_{3}^{k} - (2 d + 3 \tilde{c})_{jk}~F_{2}^{k}.
\eea
If we use (\ref{f21-5.2}) we obtain as above
$
\tilde{c} = - d,\quad \tilde{d} = c
$
so after a redefinition of the type (\ref{redef}) we obtain (\ref{ef12-n}) for
$
n = 3.
$
It is a straightforward exercise to extend, by induction, the mechanism above for all
$
n = 3,\dots,N.
$
$\qed$

We still do not have a complete basis in $V$. The reason is that the vectors
$
E_{1}^{j}, F_{1}^{j},\quad j \in J_{1}
$
might span only a subspace of 
$
V_{1}.
$
We have to add 
$
E_{1}^{j}, F_{1}^{j},\quad j \in K_{1}
$
such that
$
E_{1}^{j}, F_{1}^{j},\quad j \in J_{1} \cup K_{1}
$
is an orthonormal basis in
$
V_{1}.
$
Then we have to repeat the procedure form the preceding theorem and obtain a chain 
\be
K_{P} \subset J_{P-1} \subset \cdots K_{2} \subset K_{1}
\nonumber
\ee
and corresponding vectors 
$
\{E_{n}^{j}, F_{n}^{j}\}_{j \in K_{n}} 
$
in
$
V_{n},\quad n \geq 2
$
such that we have (\ref{EFn}). Then we determine the expressions of
$
T_{1}, T_{2}.
$
These can be easily guessed from the statement of the preceding theorem and one can repeat the 
proof to get for
$
\forall j \in K_{1}
$
\bea
T_{1}~E_{1}^{j} = E_{2}^{j}
\nonumber\\
T_{1}~F_{1}^{j} = F_{2}^{j}
\nonumber\\
T_{2}~E_{1}^{j} = - F_{2}^{j}
\nonumber\\
T_{2}~F_{1}^{j} = E_{2}^{j}
\eea
and for 
$
n = 2,\dots,P
$
and
$
\forall j \in K_{n}
$
\bea
T_{1}~E_{n}^{j} = - A_{n}^{j}~E_{n-1}^{j} + E_{n+1}^{j}
\nonumber\\
T_{1}~F_{n}^{j} = - A_{n}^{j}~F_{n-1}^{j} + F_{n+1}^{j}
\nonumber\\
T_{2}~E_{n}^{j} = - A_{n}^{j}~F_{n-1}^{j} - F_{n+1}^{j}
\nonumber\\
T_{2}~F_{n}^{j} = A_{n}^{j}~E_{n-1}^{j} + E_{n+1}^{j}.
\eea

The procedure must be repeated if 
$
E_{2}^{j}, F_{2}^{j},\quad j \in J_{2} \cup K_{2}
$
do not span the whole subspace
$
V_{2}
$.
We have to add 
$
E_{2}^{j}, F_{2}^{j},\quad j \in L_{2}
$
such that
$
E_{2}^{j}, F_{2}^{j},\quad j \in J_{2} \cup K_{2} \cup L_{2}
$
is an orthonormal basis in
$
V_{2}.
$
Then we have to repeat the procedure form the preceding theorem and obtain a chain 
\be
L_{Q} \subset L_{Q-1} \subset \cdots L_{3} \subset L_{2}
\nonumber
\ee
and corresponding vectors 
$
\{E_{n}^{j}, F_{n}^{j}\}_{j \in L_{n}} 
$
in
$
V_{n},\quad n \geq 3
$
such that we have (\ref{EFn}). The expressions of
$
T_{1}, T_{2}
$
are
$
\forall j \in L_{2}
$
\bea
T_{1}~E_{2}^{j} = E_{3}^{j}
\nonumber\\
T_{1}~F_{2}^{j} = F_{3}^{j}
\nonumber\\
T_{2}~E_{2}^{j} = - F_{3}^{j}
\nonumber\\
T_{2}~F_{2}^{j} = E_{3}^{j}
\eea
and for 
$
n = 3,\dots,Q
$
and
$
\forall j \in L_{n}
$
\bea
T_{1}~E_{n}^{j} = - A_{n}^{j}~E_{n-1}^{j} + E_{n+1}^{j}
\nonumber\\
T_{1}~F_{n}^{j} = - A_{n}^{j}~F_{n-1}^{j} + F_{n+1}^{j}
\nonumber\\
T_{2}~E_{n}^{j} = - A_{n}^{j}~F_{n-1}^{j} - F_{n+1}^{j}
\nonumber\\
T_{2}~F_{n}^{j} = A_{n}^{j}~E_{n-1}^{j} + E_{n+1}^{j}.
\eea

The procedure described above stops after a finite number of steps.

Next we determine the expression of the operator
$
T_{3}
$
on the basis obtained above and this is done using formula (\ref{f21-3.1}) and the 
expressions of
$
T_{0}, T_{1}, T_{2}
$
already obtained. We leave the derivation of these formulas as an exercise. Finally we impose 
the relations (\ref{f21-3.2}) and (\ref{f21-3.3}); in fact only one is enough, the other follows
by abstract algebra. One derives some identities which we leave, again, as an exercise. We give 
for illustration only one such consistency relation: 
\be
(1 - \alpha_{1}^{j} + 2 A_{2}^{j})~F_{1}^{j} - G_{12}~E_{1}^{j} - \cos\theta~G_{13}~E_{0}^{j} 
+ F_{1}~G_{12}~E_{0}^{j} = 0,\quad \forall j \in J_{0}.
\label{consistency}
\ee

Now we have all elements to analyze the rest of the equations (\ref{f21-2.1}) - (\ref{f21-2.9});
in fact only (\ref{f21-2.1}), (\ref{f21-2.7}), (\ref{f21-2.8}) and (\ref{f21-2.9}) should be
investigated, the others following by abstract algebra.

We make some notations:
\bea
a_{j} \equiv < E_{1}^{j}, g_{13} >, \quad b_{j} \equiv < F_{1}^{j}, g_{13} >
\nonumber\\
c_{j} \equiv < E_{2}^{j}, g_{12} >, \quad d_{j} \equiv < F_{2}^{j}, g_{12} >
\nonumber\\
\beta_{j} \equiv < E_{0}^{j}, g_{+} - g_{33} >
\eea
and
\be
\gamma \equiv \frac{m_{3}~\cos\theta}{m_{1}}.
\ee

The well-known interaction Lagrangean of the standard model assumes that there
is only one Higgs scalar field i.e.
$
|I_{3}| = 1
$
and one can derive that 
$
\gamma = 1.
$
However, in the general case we do not make this assumption and we will get new solutions, as 
explained in the Introduction. We give the results of these computations. From 
(\ref{f21-2.1}), (\ref{f21-2.7}), (\ref{f21-2.8}) and (\ref{f21-2.9}) we obtain three sets of relations:
a) in the first sector (with indices in 
$
J_{n}
$):
\bea
a_{j} = 0,\quad \forall j \in J_{1}
\nonumber\\
a_{j} = \frac{1}{2}~\gamma~\beta_{j},\quad \forall j \in J_{1}
\nonumber\\
\beta_{j} = 0,\quad \forall j \in J_{0} - J_{1}
\nonumber\\
c_{j} = 0,\quad \forall j \in J_{2}
\nonumber\\
\alpha_{1}^{j}~<E_{0}^{j}, g_{+}> = \frac{3}{4}~\gamma^{2}~\beta_{j},\quad \forall j \in J_{1} - J_{2}
\nonumber\\
\alpha_{1}^{j}~<E_{0}^{j}, g_{+}> - 2 d_{j} = \frac{3}{4}~\gamma^{2}~\beta_{j},\quad \forall j \in J_{2}
\nonumber\\
\left( 2 - \frac{1}{2}~\gamma^{2} - 2 A_{2}^{j} + 2 A_{3}^{j} - 2 m_{1}^{2}~|g_{-}|^{2} \right)~d_{j}
+ \frac{1}{2}~\gamma^{2}~A_{2}^{j}~\beta_{j} = 0,\quad \forall j \in J_{2}
\nonumber\\
\frac{1}{2}~\left( 1 + \frac{1}{2}~\gamma^{2} - \alpha_{1}^{j} + 2 A_{2}^{j} - m_{1}^{2}~|g_{13}|^{2} \right)~\beta_{j}
+ \alpha_{1}^{j}~<E_{0}^{j}, g_{33}> = 0,\quad \forall j \in J_{2}
\eea

b) in the second sector (with indices in 
$
K_{n}
$):
\bea
c_{k} = 0,\quad \forall k \in K_{1} - K_{2}
\nonumber\\
c_{k} = - \frac{3}{4}~\gamma~a_{k},\quad \forall k \in K_{2}
\nonumber\\
d_{k} = 0,\quad \forall k \in K_{1} - K_{2}
\nonumber\\
d_{k} = - \frac{3}{4}~\gamma~b_{k},\quad \forall k \in K_{2},
\nonumber\\
\left( 1 + \frac{1}{2}~\gamma^{2} + 2 A_{2}^{k} - m_{1}^{2}~|g_{13}|^{2} \right)~a_{k} = 0,
\quad \forall k \in K_{2}
\nonumber\\
\left( 1 + \frac{1}{2}~\gamma^{2} + 2 A_{2}^{k} - m_{1}^{2}~|g_{13}|^{2} \right)~b_{k} = 0,
\quad \forall k \in K_{2}
\nonumber\\
\left( 2 - \frac{1}{2}~\gamma^{2} - \frac{10}{3}~A_{2}^{k} + 2 A_{3}^{k} - m_{1}^{2}~|g_{-}|^{2} \right)~a_{k} = 0,
\quad \forall k \in K_{2}
\nonumber\\
\left( 2 - \frac{1}{2}~\gamma^{2} - \frac{10}{3}~A_{2}^{k} + 2 A_{3}^{k} - m_{1}^{2}~|g_{-}|^{2} \right) b_{k} = 0,
\quad \forall k \in K_{2}
\eea

c) in the third sector (with indices in 
$
L_{n}
$):
\bea
\left( 2 - \frac{1}{2}~\gamma^{2} + 2 A_{3}^{l} - m_{1}^{2}~|g_{-}|^{2} \right)~a_{k} = 0,
\quad \forall l \in L_{2}
\nonumber\\
\left( 2 - \frac{1}{2}~\gamma^{2} + 2 A_{3}^{l} - m_{1}^{2}~|g_{-}|^{2} \right) b_{k} = 0,
\quad \forall l \in L_{2}.
\eea

Now we use the consistency relation (\ref{consistency}) and obtain:
\be
(1 - \alpha_{1}^{j} + 2 A_{2}^{j})~\alpha_{1}^{j}~\delta_{jk} = 
\frac{3}{4}~m_{1}^{2}\gamma^{2}~\beta_{j}~\beta_{k},\quad \forall j \in J_{0}, \quad  \forall k \in J_{1}
\label{consistency1}
\ee

In particular we have:
\be
\beta_{j}~\beta_{k} = 0,\quad \forall j \in J_{0}, \quad  \forall k \in J_{1}, \quad j \not= k
\ee
so there is at most one value 
$
j_{0} \in J_{1}
$
such that
$
\beta_{j_{0}} \not= 0.
$

Finally we have all the elements to determine the solutions of our problem from Section \ref{sm}. We have two cases.
\begin{thm}
Suppose that
\be
\beta_{j} = 0,\quad \forall j \in J_{0}.
\ee
Then there are two solutions of the problem from Section \ref{sm}.

A) The space $V$ is one-dimensional, it is generated by the vector
$
E_{0} \in V_{0},\quad |E_{0}| = 1
$
and we have:
\bea
g_{11} = g_{22} = g_{33} = \frac{1}{2 m_{1}}~E_{0}
\nonumber\\
g_{ab} = 0, \quad \forall a \not= b
\eea

Moreover we have
\be
\gamma = 1 \quad \Longleftrightarrow m_{1} = m_{3}~\cos\theta
\ee
and
\be
T_{a} = 0,\quad a = 0,\dots,3.
\ee

B) The space $V$ is generated by the vector
$
E_{0} \in V_{0},\quad |E_{0}| = 1
$
and the vectors
$
E_{n}, F_{n} \in V_{n},\quad n = 2,\dots,N.
$
such that we have (\ref{EFn}). It corresponds to
$
|J_{0}| = 1,\quad J_{1} = J_{2} = \cdots = \emptyset
$
and
$
|K_{2}| = \cdots = |K_{N}| = 1
$
so we can omit the indices 
$
j, k, \dots.
$
Moreover we have:
\bea
g_{+} = g_{33} = \frac{\gamma}{2 m_{1}}~E_{0}
\nonumber\\
g_{12} = E_{2},\quad g_{-} = F_{2}
\nonumber\\
\gamma^{2} = N \qquad  \Longrightarrow \qquad  \gamma > 1
\nonumber\\
|E_{2}|^{2} = |F_{2}|^{2} \equiv \alpha_{2} = \frac{\gamma^{2} - 1}{2 m_{1}^{2}}
\nonumber\\
A_{n} \equiv \alpha_{n}/\alpha_{n-1} = \frac{n}{4}~(\gamma^{2} - n +1),\quad n = 3,\dots, N.
\eea
The expressions of the the operators
$
T_{a}, \quad a = 0,\dots,3
$
are
\be
T_{a}~E_{0} = 0, \quad a = 0,\dots,3
\ee
\bea
T_{1}~E_{2} = E_{3}
\nonumber\\
T_{1}~F_{2} = F_{3}
\nonumber\\
T_{2}~E_{2} = - F_{3}
\nonumber\\
T_{2}~F_{2} = E_{3}
\eea
and for 
$
n = 3,\dots,N
$
and
$
\forall j \in L_{n}
$
\bea
T_{1}~E_{n} = - A_{n}~E_{n-1} + E_{n+1}
\nonumber\\
T_{1}~F_{n} = - A_{n}~F_{n-1} + F_{n+1}
\nonumber\\
T_{2}~E_{n} = - A_{n}~F_{n-1} - F_{n+1}
\nonumber\\
T_{2}~F_{n} = A_{n}~E_{n-1} + E_{n+1}.
\eea
Also
\be
T_{3}~E_{n} = - \lambda_{n}~F_{n},\qquad T_{3}~F_{n} = \lambda_{n}~E_{n},\quad n = 2,\dots,N
\ee
with
\be
\lambda_{n} \equiv n~\cos\theta - \frac{\gamma^{2}}{2 \cos\theta}.
\ee
\label{AB}
\end{thm}

The other case is:
\begin{thm}
Let us suppose that 
$
\exists j_{0} \in J_{0}
$
such that
$
\beta_{j_{0}} \not= 0.
$
Then the space $V$ is generated by the vectors
$
E_{0},~E_{0}^{\prime} \in V_{0}
$
and
$
E_{1},~F_{1} \in V_{1}
$
and we have
$
\gamma < 1.
$
The norms are
\bea
|E_{0}| = 1,\quad |E_{0}^{\prime}| = \frac{1}{m_{1}^{2}~(3 \gamma^{2} + 1)}
\nonumber\\
|E_{1}|^{2} = |F_{1}|^{2} \equiv \alpha = \frac{3 \gamma^{2} + 1}{4}.
\eea
The expressions of the operators
$
T_{a}, \quad a = 0,\dots,3
$
are:
\bea
T_{1}~E_{0} = E_{1},\quad T_{2}~E_{0} = - F_{1},\quad T_{3}~E_{0} = 0
\nonumber\\
T_{1}~E_{1} = - \alpha E_{0},\quad T_{1}~F_{1} = 0, \quad T_{2}~E_{1} = 0,\quad T_{2}~F_{1} = \alpha~E_{0}
\nonumber\\
T_{3}~E_{1} = - \lambda~F_{1},\qquad T_{3}~F_{1} = \lambda~E_{1}
\eea
with
\be
\lambda = \frac{1}{\cos\theta}~\left( \sin^{2}\theta - \frac{\gamma^{2} + 1}{2} \right)
\ee
and
\be
T_{a}~E_{0}^{\prime} = 0, \quad a = 0,\dots,3
\ee
We also have:
\bea
g_{+} = \frac{3 \gamma^{2}\beta}{4 \alpha}~E_{0} + E_{0}^{\prime},\qquad
g_{33} = - \frac{\beta}{4 \alpha}~E_{0} + E_{0}^{\prime}
\nonumber\\
g_{13} = \frac{ \gamma\beta}{2 \alpha}~F_{1}, \qquad
g_{23} = \frac{ \gamma\beta}{2 \alpha}~E_{1}
\eea
with
\be
\beta \equiv \frac{3}{4 m_{1}^{2}}~\frac{(3 \gamma^{2} + 1) (1 - \gamma^{2})}{\gamma^{2}}.
\ee
\label{C}
\end{thm}

The proofs of the preceding two theorems are long but straightforward and will be omitted. Let us note that 
if we make
$
\gamma \rightarrow 1
$
in the last theorem we obtain case A from the preceding theorem. We have obtained three case corresponding to
$
\gamma = 1, \gamma > 1, \gamma < 1.
$

We still have to analyze the problem from Section \ref{sm} in the orthogonal supplement
$
V^{\bot}.
$

Let us suppose for the moment that 
$
V^{\bot} = \emptyset.
$

The first case 
$
\gamma = 1
$
corresponds then to the usual standard model. We work out the second line of the interaction Lagrangean
(\ref{T-sm}). Because we have only one Higgs field i.e.
$
|I_{3}| = 1
$
we can take
$
I_{3} = \{ H \}
$
and the non-zero expressions
$
f^{\prime}_{abc}
$
from the second line of the interaction Lagrangean (\ref{T-sm}) are:
\bea
f^{\prime}_{321} = - f'_{312} = \frac{1}{2}, \quad
f^{\prime}_{123} = - \frac{\cos~2\theta}{2\cos~\theta},\quad
f^{\prime}_{210} = \sin~\theta
\nonumber\\
f^{\prime}_{H11} = f'_{H22} = \frac{1}{2}, \quad
f^{\prime}_{H33} = \frac{1}{2\cos~\theta}
\eea
and those following from the antisymmetry property in the first two indices.
As a result we have the scalar + Yang-Mills interaction: 
\bea
T_{s + YM} = 
\sin\theta [ ( \Phi_{2}~\phi_{1\mu} - \Phi_{1}~\phi_{2\mu})~v_{0}^{\mu} 
+ m_{1}~( \Phi_{2}~\tilde{u}_{1} - \Phi_{1}~\tilde{u}_{2})~u_{0}
\nonumber\\
+ \frac{1}{2}~[ ( \Phi_{3}~\phi_{2\mu} - \Phi_{2}~\phi_{3\mu})~v_{1}^{\mu} 
+ (m_{2}~\Phi_{3}~\tilde{u}_{2} - m_{3} \Phi_{2}~\tilde{u}_{3})~u_{1}
\nonumber\\
+ ( \Phi_{1}~\phi_{3\mu} - \Phi_{3}~\phi_{1\mu})~v_{2}^{\mu} 
+ (m_{3}~\Phi_{1}~\tilde{u}_{3} - m_{1} \Phi_{3}~\tilde{u}_{1})~u_{2} ]
\nonumber\\
+ \frac{\cos2\theta}{2\cos\theta}~( \Phi_{2}~\phi_{1\mu} - \Phi_{1}~\phi_{2\mu})~v_{3}^{\mu} 
+ m_{1}~( \Phi_{2}~\tilde{u}_{1} - \Phi_{1}~\tilde{u}_{2})~u_{3}
\nonumber\\
+ \frac{1}{2}~[ ( \Phi_{H}~\phi_{1\mu} - \Phi_{1}~\partial_{\mu}\Phi_{H})~v_{1}^{\mu} 
+ m_{1}~\Phi_{H}~\tilde{u}_{1}~u_{1}
\nonumber\\
+ ( \Phi_{H}~\phi_{2\mu} - \Phi_{2}~\partial_{\mu}\Phi_{H})~v_{2}^{\mu} 
+ m_{1}~\Phi_{H}~\tilde{u}_{2}~u_{2} ]
\nonumber\\
+\frac{1}{2\cos\theta}~[ ( \Phi_{H}~\phi_{3\mu} - \Phi_{3}~\partial_{\mu}\Phi_{H})~v_{3}^{\mu} 
+ m_{3}~\Phi_{H}~\tilde{u}_{3}~u_{3} ].
\label{LagrangeanSM}
\eea

In the second case corresponding to
$
\gamma > 1
$
we can take
$
I_{3} = \{ H, H_{2}, K_{2}, \dots, H_{N}, K_{N} \}
$
so beside the Higgs field 
$
\Phi_{H}
$
there are some other (real) scalar fields
$
\Phi_{H_{n}},\Phi_{K_{n}},\quad n = 2,\dots,N.
$

The non-zero expressions
$
f^{\prime}_{abc}
$
from the second line of the interaction Lagrangean (\ref{T-sm}) are:
\bea
f^{\prime}_{321} = - f'_{312} = \frac{\gamma}{2}, \quad
f^{\prime}_{123} = - \frac{2\cos^{2}\theta - \gamma^{2}}{2cos~\theta},\quad
f^{\prime}_{210} = \sin~\theta,
\nonumber\\
f^{\prime}_{H11} = f^{\prime}_{H22} = \frac{\gamma}{2},\quad
f^{\prime}_{H33} = \frac{\gamma^{2}}{2cos~\theta}, \quad
\nonumber \\
f^{\prime}_{K_{2}11} = - f^{\prime}_{K_{2}22} = f^{\prime}_{H_{2}12} = \frac{\gamma^{2} - 1}{2m_{1}},
\nonumber \\
f^{\prime}_{H_{n+1}H_{n}1} = f^{\prime}_{K_{n+1}H_{n}1} = f^{\prime}_{H_{n+1}K_{n}2} = - f^{\prime}_{K_{n+1}H_{n}2} 
=\alpha_{n+1}
\nonumber\\
f^{\prime}_{K_{n}H_{n}3} = \frac{\gamma^{2} - 2n\cos\theta}{a\cos\theta}~\alpha_{n}
\eea
The scalar + Yang-Mills part of the Lagrangian 
$
T_{s + YM}
$
has three parts: One is generalizing the preceding
expression (\ref{LagrangeanSM})  with some minor change of the coefficients:
\bea
T_{s+YM}^{(1)} = 
\sin\theta [ ( \Phi_{2}~\phi_{1\mu} - \Phi_{1}~\phi_{2\mu})~v_{0}^{\mu} 
+ m_{1}~( \Phi_{2}~\tilde{u}_{1} - \Phi_{1}~\tilde{u}_{2})~u_{0}
\nonumber\\
+ \frac{\gamma}{2}~[ ( \Phi_{3}~\phi_{2\mu} - \Phi_{2}~\phi_{3\mu})~v_{1}^{\mu} 
+ (m_{2}~\Phi_{3}~\tilde{u}_{2} - m_{3} \Phi_{2}~\tilde{u}_{3})~u_{1}
\nonumber\\
+ ( \Phi_{1}~\phi_{3\mu} - \Phi_{3}~\phi_{1\mu})~v_{2}^{\mu} 
+ (m_{3}~\Phi_{1}~\tilde{u}_{3} - m_{1} \Phi_{3}~\tilde{u}_{1})~u_{2} ]
\nonumber\\
+ \frac{2\cos^{2}\theta - \gamma^{2}}{2\cos\theta}~[ ( \Phi_{2}~\phi_{1\mu} - \Phi_{1}~\phi_{2\mu})~v_{3}^{\mu} 
+ m_{1}~( \Phi_{2}~\tilde{u}_{1} - \Phi_{1}~\tilde{u}_{2})~u_{3} ]
\nonumber\\
+ \frac{\gamma}{2}~[ ( \Phi_{H}~\phi_{1\mu} - \Phi_{1}~\partial_{\mu}\Phi_{H})~v_{1}^{\mu} 
+ m_{1}~\Phi_{H}~\tilde{u}_{1}~u_{1}
\nonumber\\
+ ( \Phi_{H}~\phi_{2\mu} - \Phi_{2}~\partial_{\mu}\Phi_{H})~v_{2}^{\mu} 
+ m_{1}~\Phi_{H}~\tilde{u}_{2}~u_{2} ]
\nonumber\\
+\frac{\gamma^{2}}{2\cos\theta}~[ ( \Phi_{H}~\phi_{3\mu} - \Phi_{3}~\partial_{\mu}\Phi_{H})~v_{3}^{\mu} 
+ m_{3}~\Phi_{H}~\tilde{u}_{3}~u_{3} ].
\eea
The other two parts of
$
T_{s + YM}
$
contain the new scalar fields and will be not given in detail here but can be obtained from the 
expressions
$
f^{\prime}_{abc}
$
listed above.

In the third case, corresponding to
$
\gamma < 1
$
we have 
$
I_{3} = \{ H, K, H_{1}, K_{1} \}
$

The non-zero expressions
$
f^{\prime}_{abc}
$
from the second line of the interaction Lagrangean (\ref{T-sm}) are:
\bea
f^{\prime}_{321} = - f'_{312} = \frac{\gamma}{2}, \quad
f^{\prime}_{123} = - \frac{2\cos^{2}\theta - \gamma^{2}}{2cos~\theta},\quad 
f^{\prime}_{210} = \sin~\theta,
\nonumber\\
f^{\prime}_{H11} = f^{\prime}_{H22} = \frac{3\gamma^{2}\beta m_{1}}{4},\quad
f^{\prime}_{H33} = \frac{\beta m_{3}}{4 \alpha}, \quad
\nonumber \\
f^{\prime}_{K11} = f^{\prime}_{K22} = \frac{1}{m_{1}(3 \gamma^{2} + 1)},\quad
f^{\prime}_{K33} =  \frac{m_{3}}{m_{1}^{2}(3 \gamma^{2} + 1)}
\nonumber\\
f^{\prime}_{K13} = f^{\prime}_{H_{1}23} =  \frac{\gamma\beta m_{1}}{2},  \quad
f^{\prime}_{H_{1}H1} = f^{\prime}_{K_{1}H2} = \alpha, \quad
f^{\prime}_{K_{1}H_{1}3} = \lambda 
\eea
The scalar + Yang-Mills part of the Lagrangian 
$
T_{s + YM}
$
has three parts: One is generalizing the preceding
expression with some minor change of the coefficients:
\bea
T_{s+YM}^{(1)} = 
\sin\theta [ ( \Phi_{2}~\phi_{1\mu} - \Phi_{1}~\phi_{2\mu})~v_{0}^{\mu} 
+ m_{1}~( \Phi_{2}~\tilde{u}_{1} - \Phi_{1}~\tilde{u}_{2})~u_{0}
\nonumber\\
+ \frac{\gamma}{2}~[ ( \Phi_{3}~\phi_{2\mu} - \Phi_{2}~\phi_{3\mu})~v_{1}^{\mu} 
+ (m_{2}~\Phi_{3}~\tilde{u}_{2} - m_{3} \Phi_{2}~\tilde{u}_{3})~u_{1}
\nonumber\\
+ ( \Phi_{1}~\phi_{3\mu} - \Phi_{3}~\phi_{1\mu})~v_{2}^{\mu} 
+ (m_{3}~\Phi_{1}~\tilde{u}_{3} - m_{1} \Phi_{3}~\tilde{u}_{1})~u_{2} ]
\nonumber\\
+ \frac{2\cos^{2}\theta - \gamma^{2}}{2\cos\theta}~[ ( \Phi_{2}~\phi_{1\mu} - \Phi_{1}~\phi_{2\mu})~v_{3}^{\mu} 
+ m_{1}~( \Phi_{2}~\tilde{u}_{1} - \Phi_{1}~\tilde{u}_{2})~u_{3} ]
\nonumber\\
+ \frac{3\gamma^{2}\beta m_{1}}{4\alpha}~[ ( \Phi_{H}~\phi_{1\mu} - \Phi_{1}~\partial_{\mu}\Phi_{H})~v_{1}^{\mu} 
+ m_{1}~\Phi_{H}~\tilde{u}_{1}~u_{1}
\nonumber\\
+ ( \Phi_{H}~\phi_{2\mu} - \Phi_{2}~\partial_{\mu}\Phi_{H})~v_{2}^{\mu} 
+ m_{1}~\Phi_{H}~\tilde{u}_{2}~u_{2} ]
\nonumber\\
+\frac{\beta m_{3}}{4 \alpha}~[ ( \Phi_{H}~\phi_{3\mu} - \Phi_{3}~\partial_{\mu}\Phi_{H})~v_{3}^{\mu} 
+ m_{3}~\Phi_{H}~\tilde{u}_{3}~u_{3} ]
\nonumber \\
+ \frac{1}{m_{1}(3 \gamma^{2} + 1)}~[ ( \Phi_{K}~\phi_{1\mu} - \Phi_{1}~\partial_{\mu}\Phi_{K})~v_{1}^{\mu} 
+ m_{1}~\Phi_{K}~\tilde{u}_{1}~u_{1}
\nonumber\\
+ ( \Phi_{K}~\phi_{2\mu} - \Phi_{2}~\partial_{\mu}\Phi_{K})~v_{2}^{\mu} 
+ m_{2}~\Phi_{K}~\tilde{u}_{2}~u_{2} ]
\nonumber\\
+\frac{m_{3}}{m_{1}^{2}(3 \gamma^{2} + 1)}~[ ( \Phi_{K}~\phi_{3\mu} - \Phi_{3}~\partial_{\mu}\Phi_{K})~v_{3}^{\mu} 
+ m_{3}~\Phi_{K}~\tilde{u}_{3}~u_{3} ].
\eea
The other two parts of
$
T_{s + YM}
$
contain the new scalar fields and will be not given in detail here but can be obtained from the 
expressions
$
f^{\prime}_{abc}
$
listed above.
\newpage
\section{Symplectic Representations of $so(3)$\label{symplectic}}
As we have said in previous Section, in the orthogonal supplement
$
V^{\bot}
$
the operators 
$
T_{a}, a = 0,\dots,3
$
give a representation of the Lie algebra
$
u(1) \times su(2)
$;
indeed we make 
$
G_{ab} \rightarrow 0
$
in the relations (\ref{f21-1.1}) - (\ref{f21-5.3}) and we get something non-trivial only from
(\ref{f21-3.1}) - (\ref{f21-3.3}) and (\ref{f21-5.1}) - (\ref{f21-5.3}):
\bea
T_{1}~T_{2} - T_{2}~T_{1}  = \sin\theta~T_{0} + \cos\theta~T_{3}
\nonumber\\
T_{2}~T_{3} - T_{3}~T_{2}  = \cos\theta~T_{1}
\nonumber\\
T_{3}~T_{1} - T_{1}~T_{3} = \cos\theta~T_{2}
\nonumber\\
T_{0}~T_{1} - T_{1}~T_{0} = \sin\theta~T_{2}
\nonumber\\
T_{0}~T_{2} - T_{2}~T_{1} = - \sin\theta~T_{1}
\nonumber\\
T_{0}~T_{3} - T_{3}~T_{0} = 0
\label{repr}
\eea

If we define
\bea
S_{0} \equiv \cos\theta~T_{0} - \sin\theta~T_{3}
\nonumber\\
S_{1} \equiv T_{1}, \qquad S_{2} \equiv T_{2}
\nonumber\\
S_{3} \equiv \sin\theta~T_{0} + \cos\theta~T_{3}
\eea
then we obtain the standard form for a representation of
$
u(1) \times su(2)
$:
\bea
[S_{0}, S_{j} ] = 0,\quad j = 1,2,3
\nonumber\\
~[ S_{j}, S_{k} ] = \epsilon_{jkl}~S_{l}
\label{reprS}
\eea
which we will analyze in the following. We start with the second relation, i.e we determine the 
real, antisymmetric, irreducible finite dimensional representations of the algebra
$
su(2) \simeq so(3)
$.

Proceeding in analogy with theorem \ref{f012} we obtain the following result:
\begin{thm}
Let
$
S_{a}
$
be a real, antisymmetric, irreducible finite dimensional representations of the algebra
$
so(3)
$
in the real vector space $V$. Then there are three cases:

a) there exists
$
\lambda_{0} \in (0,1),~\lambda_{0} \not= 1/2
$
and the vector space $V$ can be written as
\be
V = \left(\oplus_{m=0}^{M} V_{m} \right) \oplus \left(\oplus_{n=0}^{N} W_{n} \right)
\ee
and we have the orthonormal basis
$
(E_{m}^{j},F_{m}^{j})
$
in 
$
V_{m}
$
and
$
(G_{n}^{k},H_{n}^{k})
$
in 
$
W_{n}
$
such that
\bea
S_{3}E_{m}^{j} = - (\lambda_{0} + m)~F_{m}^{j}, \qquad
S_{3}F_{m}^{j} = (\lambda_{0} + m)~F_{m}^{j},
\nonumber\\
S_{3}G_{n}^{k} = - (\lambda^{\prime}_{0} + n)~H_{n}^{k}, \qquad
S_{3}H_{n}^{k} = (\lambda^{\prime}_{0} + n)~G_{n}^{k},
\label{EF-S-a}
\eea
where
\be
\lambda^{\prime}_{0} \equiv 1 - \lambda_{0}.
\ee

b) We have
\be
V = \oplus_{m=0}^{M} V_{m}
\ee
and we have the orthonormal basis
$
E_{0}^{j}
$
in 
$
V_{0}
$
and
$
(E_{m}^{j},F_{m}^{j})
$
in 
$
V_{m}
$
such that
\bea
S_{3}E_{0}^{j} = 0,
\nonumber\\
S_{3}E_{m}^{j} = - m~F_{m}^{j}, \qquad S_{3}F_{m}^{j} = m~E_{m}^{j},
\label{EF-S-b}
\eea

c) We have
\be
V = \oplus_{m=0}^{M} V_{m}
\ee
and we have the orthonormal basis
$
(E_{m}^{j},F_{m}^{j})
$
in 
$
V_{m}
$
such that
\be
S_{3}E_{m}^{j} = -\left( \frac{1}{2} + m \right)~F_{m}^{j}, \qquad 
S_{3}F_{m}^{j} = \left( \frac{1}{2} + m \right)~E_{m}^{j}.
\label{EF-S-c}
\ee
\label{abc}
\end{thm}
{\bf Proof:} As in the proof of theorem \ref{f012}, we consider the operator
$
- (S_{3})^{2}
$
with eigenvalues $\geq 0$ and denote by
$
V_{\lambda^{2}}
$
the eigenspaces. Because of the antisymmetry of the matrix
$
S_{3}
$
it is easy to find a basis 
$
E_{j}^{\lambda}, F_{j}^{\lambda} \in V_{\lambda^{2}}
$ 
such that
\be
S_{3}E_{j}^{\lambda} = - \lambda~F_{j}^{\lambda}, \qquad
S_{3}F_{j}^{\lambda} = \lambda~F_{j}^{\lambda}
\label{EF-s}
\ee
for the values
$
\lambda > 0.
$
Then we prove, as in theorem \ref{f012} that if
$
E \in V_{\lambda^{2}}
$
then
$$
S_{1}E, S_{2}E \in V_{(\lambda + 1)^{2}} \oplus V_{(\lambda - 1)^{2}}.
$$

Let us define
$
\lambda_{0} \in [ 0, 1)
$
the minimal value for the eigenvalues 
$
\lambda.
$
The three cases from the statement correspond to: a)
$
\lambda_{0} \in ( 0, 1),~\lambda_{0} \not= \frac{1}{2} 
$;
b) 
$
\lambda_{0} = 0
$;
c) 
$
\lambda_{0} = \frac{1}{2}
$.
$\qed$

Now if we repeat the proof of theorem \ref{T12} we get after many computations that 
case a) from the preceding theorem is not possible. The cases b) and c) are described
in the following two theorems and they correspond to ``integer" and ``half-integer" spin.

\begin{thm}
In the case b) from the preceding theorem, suppose that the representation
is irreducible. Then the representations are of the type
$
{\cal D}^{M},~M = 0,1,\dots
$
acting in the the space
$
V = \oplus_{m=0}^{M} V_{m}
$
with
$
dim(V_{0}) = 1
$
and
$
dim(V_{m}) = 2,~m = 1,2,\dots,M
$
with the basis
$
E_{0} \in V_{0},~|E_{0}| = 1
$
and
$
E_{m}, F_{m} \in V_{m}
$
such that
\bea
S_{3}~E_{0} = 0
\nonumber\\
S_{3}~E_{m} = - m~\sin\theta~F_{m}, \qquad
S_{3}~F_{m} = m~\sin\theta~F_{m},\quad m = 1,2,\dots,M
\label{EFnSb}
\eea
Moreover we have
\be
S_{1}~E_{0} = E_{1}, \qquad
S_{2}~E_{0} = - F_{1}
\label{ef12-0-Sb}
\ee
\bea
S_{1}~E_{1} = - \alpha_{1}~E_{0} + E_{2}
\nonumber\\
S_{1}~F_{1} = F_{2}
\nonumber\\
S_{2}~E_{1} = - F_{2}
\nonumber\\
S_{2}~F_{1} = \alpha_{1}~E_{0} + E_{2}
\label{ef12-1-Sb}
\eea
and for
$
m = 2,\dots,M
$
\bea
S_{1}~E_{m} = - A_{m}~E_{m-1} + E_{m+1}
\nonumber\\
S_{1}~F_{m} = - A_{m}~F_{m-1} + F_{m+1}
\nonumber\\
S_{2}~E_{m} = - A_{m}~F_{m-1} - F_{m+1}
\nonumber\\
S_{2}~F_{m} = A_{m}~E_{m-1} + E_{m+1}
\label{ef12-n-Sb}
\eea
where we make the convention that 
\be
E_{M+1} = 0,\quad F_{M+1} = 0
\ee
and we have used the notations:
\be
\alpha_{m} = |E_{m}|^{2} = |F_{m}|^{2},~m = 1,\dots,M
\ee
and
\be
A_{m} = \alpha_{m}/\alpha_{m - 1},~m = 2,\dots,M.
\ee
One can explicitly determine these parameters:
\be
\alpha_{1} = \frac{1}{2}~M(M + 1)
\label{alfa-Sb}
\ee
and for 
$
m = 2,\dots,M
$
\be
A_{m} = \frac{1}{4}~(M^{2} + M - m^{2} + m) 
\label{An-Sb}
\ee

Moreover we have 
\be
S_{1}^{2} + S_{2}^{2} + S_{3}^{2} = - M(M + 1) \cdot {\bf I}.
\ee
\label{S12b}
\end{thm}

\begin{thm}
In the case c) from theorem \ref{abc}, suppose that the representation
is irreducible. Then the representations are of the type
$
{\cal D}^{M + 1/2},~M = 0,1,\dots
$
acting in the the space
$
V = \oplus_{m=0}^{M} V_{m}
$
with
$
dim(V_{m}) = 4,~m = 1,2,\dots,M
$
with the basis
$
E_{m}^{\epsilon}, F_{m}^{\epsilon},~\epsilon = \pm
$
in
$
V_{n}
$
such that
\be
S_{3}~E_{m}^{\epsilon} = - \left( m + \frac{1}{2}\right)~\sin\theta~F_{m}^{\epsilon}, \qquad
S_{3}~F_{m}^{^{\epsilon}} = ~\left( m + \frac{1}{2}\right)~\sin\theta~F_{m}^{\epsilon},\quad m = 0,1,\dots,M
\label{EFnSc}
\ee
Moreover we have
\bea
S_{1}~E_{0}^{\epsilon} =  \epsilon \frac{M+1}{2}~E_{0}^{- \epsilon} + E_{1}^{\epsilon}
\nonumber\\
S_{1}~F_{0}^{\epsilon} =  - \epsilon \frac{M+1}{2}~F_{0}^{- \epsilon} + F_{1}^{\epsilon}
\nonumber\\
S_{2}~E_{0}^{\epsilon} =  - \epsilon \frac{M+1}{2}~F_{0}^{- \epsilon} - F_{1}^{\epsilon}
\nonumber\\
S_{2}~F_{0}^{\epsilon} =  - \epsilon \frac{M+1}{2}~E_{0}^{- \epsilon} + E_{1}^{\epsilon}
\label{ef12-0-Sc}
\eea
and for
$
m = 1,\dots,M
$
\bea
S_{1}~E_{m}^{\epsilon} = - A_{m}~E_{m-1}^{\epsilon} + E_{m+1}^{\epsilon}
\nonumber\\
S_{1}~F_{m}^{\epsilon} = - A_{m}~F_{m-1}^{\epsilon} + F_{m+1}^{\epsilon}
\nonumber\\
S_{2}~E_{m}^{\epsilon} = - A_{m}~F_{m-1}^{\epsilon} - F_{m+1}^{\epsilon}
\nonumber\\
S_{2}~F_{m}^{\epsilon} = A_{m}~E_{m-1}^{\epsilon} + E_{m+1}^{\epsilon}
\label{ef12-n-Sc}
\eea
where we make the convention that 
\be
E_{M+1}^{\epsilon} = 0,\quad F_{M+1}^{\epsilon} = 0
\ee
and we have used the notations:
\be
\alpha_{m} = |E_{m}|^{2} = |F_{m}|^{2}
\ee
and
\be
A_{m} = \alpha_{m}/\alpha_{m - 1}.
\ee
One can explicitly determine these parameters:
\be
\alpha_{0} = 1
\label{alfa-Sc}
\ee
and for 
$
m = 1,\dots,M
$
\be
A_{m} = \frac{1}{4}[(M + 1)^{2} - m^{2}) 
\label{An-Sc}
\ee

Moreover we have 
\be
S_{1}^{2} + S_{2}^{2} + S_{3}^{2} = - \left(M + \frac{1}{2}\right)\left(M + \frac{3}{2}\right) \cdot {\bf I}.
\ee
\label{S12c}
\end{thm}

The proofs of these tow theorems goes along the general lines of theorem \ref{f012}. We notice
an interesting fact concerning the irreducible representations 
$
{\cal D}^{M/2 + 1},~M = 0,1,\dots
$
from the preceding theorem. It is known that Schur lemma is not valid for real irreducible
representations: according to \cite{V}, pg. 153 this lemma is valid only on vector spaces
over a division field which is algebraically closed. This is not the case for the real numbers
and indeed we can find a non-trivial operator commuting with this representations.
\bea
S_{0}~E_{n}^{+} = - \lambda~F_{n}^{+}\qquad
S_{0}~F_{n}^{+} = \lambda~E_{n}^{+}
\nonumber\\
S_{0}~E_{n}^{-} = \lambda~F_{n}^{-}\qquad
S_{0}~F_{n}^{-} = - \lambda~E_{n}^{-}
\label{s0}
\eea
for any
$
\lambda > 0.
$

Finally we extend the results obtained above for representations of the algebra
$
S_{a},~a = 0,\dots,3
$
from (\ref{reprS}). Suppose that we have an real, antisymmetric, irreducible representation of this 
algebra. The operators
$
S_{a},~a = 1,2,3
$
given then a real, antisymmetric, but not necessarily irreducible representation of the 
$
so(3)
$
algebra so it must fall in the cases b) and c) of theorem \ref{abc}. In the case b) one can 
prove rather easily that there are two types of irreducible representations:

1) Of the type
$
{\cal D}^{M}
$
with
$
S_{0} = 0
$;

2) Of the type
$
{\cal D}^{M} \oplus {\cal D}^{M}
$
with the basis
$
E_{0}^{\epsilon} \in V_{0}
$
and
$
E_{m}^{\epsilon}, F_{m}^{\epsilon} \in V_{m}~(m = 1,\dots,M)
$
and the actions of the representation is for 
$
S_{0}
$
given by
\bea
S_{0}~E_{0}^{\epsilon} = - \epsilon \mu~E_{0}^{- \epsilon}
\nonumber\\
S_{0}~E_{n}^{\epsilon} = ~\epsilon \mu E_{n}^{ - \epsilon},\qquad
S_{0}~F_{n}^{\epsilon} = ~\epsilon \mu F_{n}^{ - \epsilon}
\eea
and for 
$
S_{a},~a = 1,\dots,3
$
is given by the expressions from theorem \ref{S12b} for both values of
$
\epsilon.
$

3) In case c) the irreducible representations are
$
{\cal D}^{M/2 + 1},~M = 0,1,\dots
$
extended by (\ref{s0}).

So, the general solution of the problem from Section \ref{sm} is given by a direct sum between one of
the three solutions from theorems \ref{AB} and \ref{C} and some direct sum of representations of type
1), 2) and 3) described above. A priori there is no way to discriminate between them. However, such a criterion
is given by the super-renormalizablility condition recently proposed in \cite{sr3}. After some tedious computations
one can prove that only the first solution from theorem \ref{AB} is admissible i.e. the usual solution of the 
standard model, but we have to add some real and antisymmetric representation 
$
S_{a},~a = 0,\dots,3
$ 
of the Lie algebra
$
su(1) \times so(3)
$
and this means that we must add to the Lagrangean (\ref{LagrangeanSM}) an extra piece of the form
\bea
T = (S_{a})_{jk} \Phi_{j}~\partial_{\mu}\Phi_{k}~v_{a}^{\mu}.
\label{S}
\eea
This means that there are some extra Higgs fields in the nature, beside the usual one. The solution for 
the representation 
$
S_{a}.~a = 0,\dots,3
$ 
is not unique.
\section{Conclusions}

We have seen that gauge invariance in the first two orders of the perturbation theory for Yang-Mills
models leads to a problem of classification of symplectic representations i.e. real, antisymmetric (and
irreducible) representations of the Lie algebra relevant for the model; in our case
$
u(1) \otimes su(2) \simeq u(1) \otimes so(3)
$.

The theory of real representations of real Lie algebras is known in the literature \cite{Iwa}, 
\cite{Oni} but we think that the particular case of antisymmetric representations deserves
a special treatment in analogy with the complex case and we provided this here for the algebra
considered above.

From the physical point of view it follows from above that in the case of a single Higgs field i.e.
$
|I_{3}| = 1
$
we get exactly the Yang-Mills Lagrangean of the standard model. In the general case we have much more
solutions and a way to select the physical ones is to use the condition of super-renormalizablility.
This condition reduces the arbitrariness of this representation but does not fix it uniquely. 

In forthcoming papers we will investigate if this condition can be implemented
for some extensions of the standard model. Also it is interesting to see if the condition of super-renormalizablility
applied to higher orders of perturbation theory reduces even further the arbitrariness of $S$ from (\ref{S}).

\end{document}